\documentclass[12pt]{article}

\usepackage{amsfonts,amssymb,amsmath}

\newcommand{\ch}{\cosh}
\newcommand{\sh}{\sinh}
\newcommand\ptl{\partial}
\newcommand\br{\langle}
\newcommand\kt{\rangle}
\newcommand\upa{{\uparrow}}
\newcommand\dna{{\downarrow}}
\newcommand\vep{\varepsilon}
\newcommand\Hub{\mathrm{Hub}}

\begin{document}

\pagestyle{empty}

\setcounter{page}{0}

\hspace{-1cm}


\begin{center}

{\LARGE{\sffamily  Supergroup approach to the Hubbard model}}\\[1cm]

{\large V.Kirchanov, V. Zharkov \footnote{vita@psu.ru,  Kirchanv@rambler.ru}\\[.484cm] 
\textit{ Perm State Technical university, Komsomolcky Prospect, 29a, Perm, 614600, Russia\\[.242cm]
Natural Sciences Institute of Perm State university,\\
 Genkel st.4, Perm,614990, Russia. }}
\end{center}
\vspace{2.5em}

\begin{abstract}
Based on the revealed hidden supergroup structure, we develop a new approach
to the Hubbard model. We reveal a relation of even Hubbard operators to the
spinor representation of the generators of the rotation group of
four-dimensional spaces. We propose a procedure for constructing a matrix
representation of translation generators, yielding a curved space on which
dynamic superfields are defined. We construct a new deformed nonlinear
superalgebra for the regime of spinless Hubbard model fermions in the case
of large on-site repulsion and evaluate the effective functional for
spinless fermions.
\end{abstract}

\newpage
\pagestyle{plain}

\section{Introduction}

\noindent

The Hubbard model~[1] remains the main test ground for investigating the
effects of strong correlations between electrons. Traditionally, the effects
of a strong Coulomb interaction play an important role in understanding the
mechanisms of high-temperature superconductivity, the physics of the
Mott--Hubbard-type metal--insulator transition, and the related magnetic
states. Experimental evidence of a spin liquid in organic metal
(BEDT--TTF)$_2$Cu$_2$(CN)$_3$~[2] additionally complicated the situation: it
has become necessary to take the spin liquid into account in this model. We
believe that this has revitalized the problem of developing new approaches
to the Hubbard model based on field theory methods and especially on using
the functional integral techniques.

Using the path integral proposed for the Hubbard model in~[3],~[4], we here
continue developing the superfield formulation. We discover a nonlinear
hidden supergroup occurring in strongly correlated systems and show how a
matrix representation for the coordinate shift generators can be introduced
together with the collective dynamic fields. We compute the effective
functional based on the new nonlinear superalgebra, also obtained here.

We recall the crucial points of the approach in~[3]. As the starting
formulation, we take the atomic representation of the Hubbard model. A
property of this description is that it allows introducing a local
supergroup whose generators are defined by the full set of the on-site
Hubbard operators. The supergroup generators act globally and are
independent of the lattice coordinates. Next, a supercoherent state is
constructed that specifies a superorbit of our supergroup and contains a set
of dynamic (i.e., time and coordinate dependent) fermionic and bosonic
fields. Passing from the operator to the superfield formulation is achieved
using the effective functional of the form
\begin{equation}
\hat L=\frac{\br G|\ptl/\ptl\tau-H_\Hub|G\kt}{\br G|G\kt},
\end{equation}
where $|G\kt$ is the supercoherent state that specifies the supergroup orbit
to be defined below and $H_\Hub$ is an operator expression for the Hubbard
model.

We emphasize that one of the fundamental questions of strongly correlated
systems is the problem of the existence of a hidden supergroup structure and
the procedure for introducing it into the Hubbard model. Identifying this
superstructure would allow clarifying the question of new symmetries and
supersymmetries and developing a procedure for calculating effective
superfield functionals for the strong Coulomb interaction systems described
by models that are nonlinear in the dynamic bosonic and fermionic fields.

A central element of the proposed approach is the procedure for identifying
even Hubbard operators with local generators of the Lorentz group and other
rotation groups of four-dimensional spaces. In strongly correlated systems,
we understand the Lorentz group to be a four-dimensional rotation group
containing hyperbolic rotations with a ``speed of light" constant depending
on the characteristic energy, for example, the parameter of the electron
band width or the exchange integral. The full set of Hubbard operators is
identified with a Lorentz group superextension, which is a subgroup in the
superconformal group.

Another important point relates to the introduction of a matrix
representation for the group of translations. This subgroup, via a quotient,
allows introducing the coordinate space on which the dynamic bosonic and
fermionic fields are defined. We introduce the subgroup of translations
using the procedure not of the group affinization, as in the majority of
papers (see, e.g.,~[5]), but of the deformation and contraction of algebras.
Our approach is more complicated but seems to be the only one suitable for
the Hubbard model. The tricks that we use allow evaluating the effective
functional for some deformation of the algebra of spinless fermions, which
is nonlinear in the Fermi and Bose generators.

We show how the hidden supergroup structure emerges in the Hubbard model.
Any quantum system is characterized by the wave function $\Psi$, which is a
function of the coordinates $x$, $y$, and $z$ and time $t$, i.e., $\Psi=
\Psi(x,y,z,t)$. The most widespread groups act on this wave function via
operators defined in differential form. For example, one of the widespread
continuous groups acting on the wave function is the Poincar\'e group, which
contains the space--time translation operators
$$
P_x=-i\frac{\ptl}{\ptl x},\qquad
P_y=-i\frac{\ptl}{\ptl y},\qquad
P_z=-i\frac{\ptl}{\ptl z},\qquad
P_t=i\frac1c\frac{\ptl}{\ptl t},
$$
the operators of spatial rotations in the $(x^i,x^j)$ planes
$$
L_z=i\biggl(y\frac{\ptl}{\ptl x}-x\frac{\ptl}{\ptl y}\biggr),\qquad
L_y=i\biggl(x\frac{\ptl}{\ptl z}-z\frac{\ptl}{\ptl x}\biggr),\qquad
L_x=i\biggl(z\frac{\ptl}{\ptl y}-y\frac{\ptl}{\ptl z}\biggr),
$$
and the operators of hyperbolic rotations (boosts) in the$(ct,x^k)$ planes
$$
L_1=i\biggl(t\frac{\ptl}{\ptl x}+x\frac{\ptl}{\ptl t}\biggr),\qquad
L_2=i\biggl(t\frac{\ptl}{\ptl y}+y\frac{\ptl}{\ptl t}\biggr),\qquad
L_3=i\biggl(t\frac{\ptl}{\ptl z}+z\frac{\ptl}{\ptl t}\biggr).
$$
The action of a finite group on the wave function is given by the
exponential map and has the form
\begin{equation}
e^{E_iP_i+H_jL_j+H_k\Sigma_k}\Psi(x,y,z,t)=e^{H_k\Sigma_k}\Psi(Ax+B),
\end{equation}
where $E_i$ and $H_k$ are small parameters, summation over the indices
$i,j,k=1,2,3$ is understood, the matrix $A$ defines a vector representation
of the rotation group, and the matrix $B$ represents shifts. We note that
the parameters $H_k$ determine both the matrix $A$ acting on the coordinates
and the elements of the rotation group for the function $\Psi $ in the
spinor basis. The wave function $\Psi $ is defined in the spinor
representation and is acted upon by elements of the group parameterized by
the $H_k$.

\section{Hubbard model}

The Hubbard model~[1],~[6] is one of the fundamental models describing
systems of strongly interacting electrons in solids. It underlies the
description of the band magnetism and superconductivity in strongly
correlated metals and the metal--insulator phase transition in solids.

The Hubbard model involves electrons that are nondegenerate with respect to
the orbital state and propagate by hops over sites of the crystal lattice
with a Coulomb interaction at one site. In the secondary quantization
representation, the Hubbard model Hamiltonian has the form
\begin{equation}
\widehat H=-W\sum_{i,j,\sigma}{a_{i\sigma}^+}a_{j\sigma}+
U\sum_i{n_{i\upa}}n_{i\dna}-\mu\sum_{i,\sigma}{n_{i\sigma}},
\end{equation}
where $a_{i\sigma}^+ $ and $a_{i\sigma}$ are Fermi operators of creation and
annihilation of an electron with the spin $\sigma$ at the $i$th site,
$n_{i\sigma}=a_{i\sigma}^+a_{i\sigma}$ is the operator of the number of
electrons with a given spin $\sigma$ at the $i$th site, and the spin
$\sigma$ takes two values $\upa$ $(+)$ and $\dna$ $(-)$.

The Hubbard model involves only three parameters: the matrix element $W$ of
the electron transition to the neighboring site, which is related to the
electron band width, the parameter $U$ of Coulomb repulsion of two electrons
at a site $i$, and the chemical potential $\mu $ or the electron
concentration $n$ (the average number of electrons per lattice site).

In what follows, we set $W=1$ and consider the half-filled band in the case
of strong repulsion. In that regime, the lowest-lying states are described
by magnetic fields together with fermionic degrees of freedom. The repulsion
parameter practically drops out of the problem.

We are interested in the two-dimensional Hubbard model, and we start
investigating it from the atomic limit.

\section{Hubbard operators}

Under the condition of a strong Coulomb repulsion, the Coulomb term in the
Hamiltonian is taken as the zeroth approximation. Then the zeroth
approximation reduces to a one-site problem and can be solved in the basis
of localized atomic functions:
\begin{equation}
|ip\kt\colon|i0\kt,\;|i+\kt,\;|i-\kt,\;|i2\kt,
\end{equation}
where $|i0\kt$ is the state in which the $i$th site does not contain an
electron, $|i+\kt\equiv|i\upa\kt$ is the state with one spin-up electron at the $i$th
site, $|i-\kt\equiv|i\dna\kt$ is the state with one spin-down electron at the $i$th
site, and $|i2\kt=|i\upa\dna\kt$ is the state with two electrons at the
$i$th site, one with spin up and the other with spin down. Any state at the
site $i$ can be represented as a superposition of these localized functions:
\begin{equation}
\psi_i=\alpha_i|i0\kt+\beta_i|i+\kt+\gamma_i|i-\kt+\delta_i|i2\kt,
\end{equation}
where $\alpha_i$, $\beta_i$, $\gamma_i$, and $\delta_i$ are coefficients of
the atomic basis functions with $\beta_i$ and $\gamma_i$ being Grassmann
numbers. In what follows, we use matrix elements of the superspinor
representation of the local supergroup to represent the coefficients of this
decomposition.

Transitions between the atomic basis states are described by 4$\times$4
matrices corresponding to the {\sl Hubbard operators}:
\begin{equation}
X_i^{pq}=|ip\kt\br iq|=\begin{pmatrix}
{X_i^{00}}&{X_i^{0+}}&{X_i^{0-}}&{X_i^{02}}\\[1mm]
{X_i^{+0}}&{X_i^{++}}&{X_i^{+-}}&{X_i^{+2}}\\[1mm]
{X_i^{-0}}&{X_i^{-+}}&{X_i^{--}}&{X_i^{-2}}\\[1mm]
{X_i^{20}}&{X_i^{2+}}&{X_i^{2-}}&{X_i^{22}}\end{pmatrix}.
\end{equation}
Each Hubbard operator corresponds to a matrix with the unity at the
intersection of the $p$th row and the $q$th column and with all other
elements equal to zero. The Hubbard operators satisfy the relations
$X_i^{pq}X_i^{rs}=\delta_{qr}X_i^{ps}$, which define their algebra, and the
completeness condition $\sum_pX_i^{pp}=4$ for the diagonal elements.

The Fermi (f-type) operators are $X_i^{0+}$, $X_i^{0-}$, $X_i^{+2}$,
$X_i^{-2}$, $X_i^{+0}$, $X_i^{-0}$, $X_i^{2+}$, and $X_i^{2-}$ because they
change the number of electrons at a site by an odd number, 1 or 3. They
satisfy the anticommutation relations
\begin{equation}
[X_i^{pq},X_j^{rs}]_+=X_i^{pq}X_j^{rs}+X_j^{rs}X_i^{pq}=
\delta_{ij}(\delta_{qr}X_i^{ps}+\delta_{qr}X_i^{rq}).
\end{equation}
The Bose (b-type) operators are $X_i^{+-}$, $X_i^{-+}$, $X_i^{20}$, and
$X_i^{02}$. They change the number of electrons at a site by an even number,
0 or 2. They satisfy the commutation relations
\begin{equation}
[X_i^{pq},X_j^{rs}]_-=X_i^{pq}X_j^{rs}-X_j^{rs}X_i^{pq}=
\delta_{ij}(\delta_{qr}X_i^{ps}-\delta_{sp}X_i^{rq}).
\end{equation}
This formula also holds if it involves at least one b-type operator or one
of the diagonal operators $X_i^{00}$, $X_i^{++}$, $X_i^{--}$, or $X_i^{22}$.

In terms of the $X$ operator, the Hubbard model Hamiltonian has the form
\begin{align}
H={}&\sum_i\{-\mu X_i^{++}-\mu X_i^{--}+(U-2\mu)X_i^{22}\}-{}
\nonumber
\\[2mm]
&{}-W\sum_{ij}\{(X_i^{0+}+X_i^{2-})(X_j^{+0}+X_j^{-2})+
(X_i^{-0}+X_i^{2+})(X_j^{0-}+X_j^{+2})\},
\end{align}
where the first sum also includes the chemical potential $\mu$. In this
representation, the Coulomb term is linear, and the kinetic energy term is
given by a bilinear form in the $X$ operators. In what follows, we use this
operator formulation because it takes us precisely to the supergroup
construction, which, in our opinion, is the central element of all strongly
correlated models.

\section{Minimal superalgebra in the Hubbard model}

We use the construction of a ``symmetry tower" introduced for the Hubbard
model in~[4]. We consider the regime of a half-filled band at strong
repulsion such that the energy of the $|+\kt$ and $|-\kt$ states is lower
than the energy of the $|0\kt$ and $|2\kt$ states by $U/2$. Under strong
repulsion, it is reasonable to proceed from the so-called spinless
superalgebra, which is constructed as follows.

1. We start with the superalgebra of Hubbard operators, denoted by $S2$.
From the eight Fermi-like Hubbard operators, we form creation and
annihilation operators taking two orientations of the spin $\sigma$ into
account:
\begin{equation}
\begin{alignedat}2
&a_\upa=X^{0+}+X^{-2},&\qquad&a_\upa^+=X^{+0}+X^{2-},
\\[2mm]
&a_\dna=X^{0-}-X^{+2},&\qquad&a_\dna^+=X^{-0}-X^{2-}.
\end{alignedat}
\end{equation}
Their anticommutation relations are
\begin{equation}
\{a_\sigma^+,a_{\sigma'}\}_+=a_\sigma^+a_{\sigma'}+
a_{\sigma'}a_\sigma^+=\delta_{\sigma\sigma'}.
\end{equation}
These operators differ from the Hubbard operators in their (anti)commutation
relations and therefore give a new algebra, which together with the density
and spin operators describes excitations in the metallic state. In~[4], this
algebra was denoted by $S1$. It differs from the algebra $S2$ of Hubbard
operators.

2. We introduce spinless operators $a$ and $a^+$ as
\begin{equation}
a=\frac{a_\upa+a_\dna}{\sqrt2},\qquad
a^+=\frac{a_\upa^++a_\dna^+}{\sqrt2},\qquad\rho=\frac12a^+a.
\end{equation}
The operators $(a^+,a,\rho)$ constitute an algebra, denoted here by $S0$,
because we have the (anti)commutation relations
\begin{equation}
\{a,a^+\}_+=\hat1,\qquad[\rho ,a^+]=a^+,\qquad[\rho,a]=-a.
\end{equation}
In a matrix representation of this algebra, we construct a product of the
set of dynamic Bose and Fermi fields with the corresponding generators. Just
this product enters the nonlinear representation of the supergroup via the
exponential map of the form
\begin{equation}
U=\exp\begin{pmatrix}-E_z&\chi&\chi&0\\[1mm]\chi^*&H_z&H^+&-\chi\\[1mm]
\chi^*&H^-&-H_z&\chi\\[1mm]0&-\chi^*&\chi^*&E_z\end{pmatrix},
\end{equation}
where $E_z$ and $H^\pm$ are field variables and $\chi$ and $\chi^*$ are
Grassmann variables.

3. We now form a spin vector $\vec{s}=(s^+,s^-,s_z)$ from the operators
introduced in item~1:
\begin{equation}
s^+=\frac1{\sqrt2}\,a_\upa^+a_\dna,\qquad
s^-=\frac1{\sqrt2}\,a_\dna^+a_\upa,\qquad
s_z=\frac12(a_\upa^+a_\upa-a_\dna^+a_\dna).
\end{equation}
The commutation relations for the spin operators are
\begin{equation}
[s^+,s^-]_-=s_z,\qquad[s_z,s^+]_-=s^+,\qquad[s_z,s^-]_-=-s^-.
\end{equation}
These operators constitute a spin algebra for the spin $1/2$.

The spin operators are given by a 2$\times$2 matrix; in the
four-dimensional square matrix in~(14), the product of the magnetic field
vector times the spin vector yields a matrix of the form
$$
\begin{pmatrix}0&0&0&0\\[1mm]0&H_z&H^+&0\\[1mm]
0&H^-&-H_z&0\\[1mm]0&0&0&0\end{pmatrix}.
$$

4. From the operators introduced in item~1, we also form a vector
$\vec\rho=(\rho^+,\rho^-,\rho_z)$ of the density operator:
\begin{equation}
\rho^+=\frac1{\sqrt2}a_\upa^+a_\dna^+,\qquad
\rho^-=\frac1{\sqrt2}a_\dna a_\upa,\qquad
\rho_z=\frac12(a_\upa a_\upa^+-a_\dna^+a_\dna).
\end{equation}
These operators constitute an algebra with the commutation relations
\begin{equation}
[\rho^+,\rho^-]_-=\rho_z,\qquad[\rho_z,\rho^+]_-=\rho^+,\qquad
[\rho_z,\rho^-]_-=-\rho^-.
\end{equation}
In the matrix representation, this algebra yields the matrix of density
fields
$$
\begin{pmatrix}E_z&0&0&E^+\\[1mm]0&0&0&0\\[1mm]
0&0&0&0\\[1mm]E^-&0&0&-E_z\end{pmatrix}.
$$
A second algebra of density operators with the operator components
$\vec\rho=(a_\upa a_\dna,a_\dna^+a_\upa^+,a_\dna a_\dna^+-a_\upa^+a_\upa)$
and the same commutation relations is also possible.

We thus formed three algebras from the Fermi-like Hubbard operators: one
fermionic spinless and two bosonic algebras. The fermionic algebra describes
creation and annihilation of spinless electrons at the sites, the spin
algebra is used for taking the interaction of the magnetic moments of
electrons into account, and the density bosonic algebra is designed for
describing the interaction of density fluctuations, for example, for the
analysis of density charge waves.

\subsection{Deformation and contraction of a Lie algebra}
Contraction of a Lie algebra is a limit operation of a Lie algebra
deformation. Because these techniques have not been previously used in the
Hubbard model, we give the necessary definitions below, referring
to~[7]--[10].

Let $G$ be the Lorentz group, $g$ be the Lie algebra of the Lorentz group,
and $a$ be the subalgebra of the rotation group of the three-dimensional
space. Then a contraction of $g$ gives the Lie algebra of the inhomogeneous
rotation group. Indeed, let $\{c_{ij}^k\}$ be the set of structure constants
of the Lie algebra $g$ in a fixed basis $e_1,\dots,e_n$, and let $A(t)$ be a
curve in the group of nondegenerate linear transformations of the $g$ group
space such that $A(1)=E$. Let $e_i(t)=A(t)e_i$ and $c_{ij}^k(t)$ be
structure constants of $g$ in the basis $\{e_i(t)\}$. If the structure
constants change as $t$ varies, then the algebra is said to be deformed. If
some of the structure constants vanish as $t$ tends to zero or infinity,
then such a process of changing the algebra is called a
contraction~[7]--[10].

\subsection{Local supergroup in the Hubbard model}
The Hubbard operators depend on the index parameterizing the coordinate,
i.e., they are given by direct products of local copies of a set of
generators defined at a selected site. We note that this set of
coordinate-independent generators constitutes a global superalgebra (i.e.,
one defined on the entire space). It can be used to construct the local
supergroup in the Hubbard model by an exponential map. The dynamics of the
system is then given by the dynamic Fermi and Bose fields in the exponential
representation. We consider the spinless case in what follows.

\section{Even Hubbard operators as generators of the group
of four-dimensional rotations}

We introduce two triples of operators: $(X^{00}-X^{22},X^{02},X^{20})$ and
$(X^{++}-X^{--},X^{+-},X^{-+})$. The commutators of these operators are
expressed only in terms of operators from the same triple. This means that
the operators divide into two groups, which commute with each other, i.e.,
define either a direct product of two $SU(2)$ groups or the direct product
$SU(2)\otimes SU(1,1)$ of two subgroups. Naturally, the choice of one of
these two types of groups must correspond to a certain choice of the fields.

We note that if the commutation relations in the second subalgebra are
chosen such that it becomes the $su(1,1)$ algebra, then this leads to a
factor $i$ appearing in front of some of the fields in the functional
integral (a Wick rotation), and the group and the space hence become
hyperbolic.

If linear combinations (half-sum or half-difference) of the generators of
these two groups are taken, then it can be easily verified that their
commutation relations coincide with the commutation relations of the
four-dimensional Lorentz group~[11]. Depending on whether hyperbolic
rotations are introduced, this rotation group can be made into the group of
rotations of the four-dimensional Euclidean space. We obtain the structure
of the commutation relations for the six generators:
\begin{equation}
[L_i,L_j]=i\vep_{ijk}L_k,\qquad[L_i,K_j]=i\vep_{ijk}K_k,\qquad
[K_i,K_j]=i\vep_{ijk}L_k.
\end{equation}
Here, $L_i$, $i=1,2,3$, specify the spatial rotation generators, $K_j$,
$j=1,2,3$, are the generators of the time axis rotations, and $\vep_{ijk}$
is the Levi-Civita tensor. Each of these groups constructed on the given
generators is a three-parameter group. Three parameters, multiplication by
the rotation generators in the exponential map, are responsible for the
rotations, and three are responsible for the hyperbolic rotations (boosts).
Hubbard operators are proportional to the following combinations: the spin
operators $s_i=L_i+K_i $, and the density operators $\rho_i=L_i-K_i $,
$i=1,2,3$.

We hence conclude that the Bose Hubbard operators can be divided into two
subgroups, from which the full set of the Lorentz group generators in a
spinor basis can be constructed.

We note that the Lorentz group is a subgroup of the conformal group, which
is the largest nonlinear group of the four-dimensional space--time on which
the wave function of the system is defined. Thus extending the group of
rotations of the four-dimensional space to the conformal and further to the
superconformal group, we can construct a local (super)group of the Hubbard
model. Because the atomic basis is a central point in practically all of the
strongly interacting systems, we have revealed a hidden symmetry in strongly
correlated system and indicated a way to introduce local (super)groups for
the popular models.

Another important point refers to the introduction of a matrix
representation for shift generators. The shift generators are usually given
by differential operators (derivatives). In the Hubbard model, just a matrix
representation is necessary, different from the one that follows from the
affinization procedure~[5]. We introduce shift operators in terms of
decreasing the dimension of the rotation group and the procedure of
deforming the algebra of the Lorentz group generators in the
four-dimensional space. This procedure is more complicated than
affinization, but we believe it is the only one applicable to the Hubbard
model.

We introduce a deformation procedure and then a contraction of the even
Hubbard generators. We multiply the generators of hyperbolic rotations by a
constant $R$, $K_i'=RK_i$, $i=1,2,3$, such that the commutation relations
become
\begin{equation}
[K'_i,K'_j]=i\biggl(\frac1{R^2}\biggr)\vep_{ijk}L_k.
\end{equation}
In what follows, similarly to~[12], we regard $L_i$ as the generators of
rotations in the three-dimensional space and $K'_i$ as the generators of
translations. It follows from~(20) that via such a transition from the
four-dimensional to the three-dimensional space, we have effected a
transition from the group of four-dimensional rotations to the group of
inhomogeneous three-dimensional rotations, which contain translations in a
curved space. In the limit as $R\to\infty$, we obtain the three-dimensional
Poincar\'e group $ISO(3)$, with the algebra of generators
\begin{equation}
[L_i,L_j]=i\vep_{ijk}L_k,\qquad[L_i,P_j]=i\vep_{ijk}P_k,\qquad[P_i ,P_j]=0,
\end{equation}
where $P_i=K_j'$ are generators of translations in flat space. The boost
generators thus define translations in a curved space whose radius is
defined by the expression $E^2=E_z^2+E^+E^-$, and their identification with
the generators of translations in the curved space allows introducing the
coordinates of the base space by replacing $(E^+,E^-,E_z)$ with $(x,y,z)$.
We see in what follows that this expression naturally enters the matrix
elements of the supergroup and determines the radius of the space where
magnetic dynamic fluctuations occur.

\section{The group space}

The group structure that we have revealed on the even Hubbard generators has
a subgroup and a quotient group. Based on this, we can introduce the group
of motions specifying the coordinate functions and the functions that are
dynamic fields in our problem~[3]. We consider the finite continuous group
$G$ depending on the parameters $E_z$, $E^+$, $E^-$ and $H_z$, $H^+$, $H^-$.
In what follows, we use the notation $E_z$, $E^+$, $E^-$ with the
understanding that these can be replaced with $(x,y,z)$ whenever necessary.
The parameters $\{a,b\}=\{E_z,E^+,E^-;H_z,H^+,H^-\}$ can be considered the
coordinates of a point in the $(3{+}3)$-dimensional group space. Each point
of the space is set in correspondence with a transformation from the group
$G$. The point corresponding to the identity transformation is called the
initial point of the space. The initial and an arbitrary point of the space
define a vector. Any infinitesimal transformation is expressed in terms of
the generators of the quotient group $X_k$ and the subgroup $Y_\alpha$:
\begin{equation}
dG=i(da^kX_k+db^\alpha Y_\alpha),
\end{equation}
where $a^k$ and $b^\alpha $ are parameters of the group, $k=1,2,3$, and
$\alpha=1,2,3$. We again note that the quotient group here specifies the
``true" coordinates $(x,y,z)$.

The 1-forms are expressed in terms of the group elements as
\begin{equation}
G\,dG^{-1}=G^{-1}\,dG=i(\omega^iX_i+\theta^\alpha Y_\alpha),
\end{equation}
where $X_i$ is a generator of the quotient group~[13].

\section{Structure equations for the group space}

Structure equations for the group space coincide in form with the
Maurer--Cartan structure equations for a Riemannian space with zero torsion
and a nonzero curvature~[14]--[17]:
\begin{equation}
\begin{aligned}
&d\omega^i=[\omega^k\omega_k^i]\equiv\omega^k\wedge\omega_k^i,
\\[2mm]
&d\omega_i^j=[\omega_i^k\omega_k^j]+R_{i[kh]}^j[\omega^k\omega^h]\equiv
\omega_i^k\wedge\omega_k^i+R_{i[kh]}^j\omega^k\wedge\omega^h.
\end{aligned}
\end{equation}
The forms $\omega^i$ are components of an infinitesimal shift of the origin
of a frame with respect to the frame at a point $a$. The forms $\omega_j^i$
are a change in the components of the frame itself. A transformation from
the group $G$ is a rotation if it belongs to the subgroup $H$. The
transformations from the subgroup $H$ leave the origin of the group space
fixed and constitute the so-called stationary subgroup of this space. A
transformation is a shift if it is generated by an infinitesimal
transformation $\omega^iX_i$.

A general transformation of the group $G$ is represented as the product
\begin{equation}
G=K(a)H(b),
\end{equation}
where $K(a)=e^{ia^jX_j}$ is a transformation belonging to the left coset
class $G/H$ of the group $G$ with respect to the subgroup $H$ and
$H(b)=e^{ib^\alpha Y_\alpha}$ is a transformation belonging to the subgroup
$H$.

\section{Cartan forms for the Bose Hubbard operators}

We consider vectors $(x,y,z)$ of the coordinates of the $SO(3)$ or $SO(2,1)$
group space, which in matrix~(14) are expressed as components of the
electric field $E_z$, $E^+$, $E^-$, the magnetic field $H_z(x,y,z)$,
$H^+(x,y,z)$, $H^-(x,y,z)$, and the Grassmann $\chi(x,y,z)$ and conjugate
$\chi^*(x,y,z)$ fermionic fields as functions of the space--time coordinates
$(x,y,z)$. We use the local spherical coordinates $(E,\theta ,\varphi)$ and
$(H,\sigma,\vartheta)$ to write the electric and magnetic field vectors:
\begin{alignat*}3
&E_z=E\cos\theta,&\qquad&E^+=E\sin\theta\cdot e^{i\varphi},&\qquad
&E^-=E\sin\theta\cdot e^{-i\varphi},
\\[2mm]
&H_z=H\cos\sigma,&\qquad&H^+=H\sin\sigma\cdot e^{i\vartheta},&\qquad
&H^-=H\sin\sigma\cdot e^{-i\vartheta}.
\end{alignat*}
Then
\begin{equation}
E^2=E_z^2+E^+ E^-,\qquad H^2=H_z^2+H^+ H^-.
\end{equation}
We place these fields into a supermatrix and write the exponential map in
the form
\begin{equation}
U=\exp\begin{pmatrix}E_z&\chi_1&\chi_2&E^+\\[1mm]
\chi_1^*&H_z&H^+&\chi_3\\[1mm]\chi_2^*&H^-&-H_z&\chi_4\\[1mm]
E^-&\chi_3^*&\chi_4^*&-E_z\end{pmatrix}.
\end{equation}
Here and hereafter, in the expression for the exponential map, we do not
write the standard minus sign in front of the matrix in the exponent. This
sign can easily be compensated at the final stage by redefining the fields.
As a result, we obtain a nonlinear representation for the supergroup that
locally performs superrotations of the atomic basis. The problem of
calculating the matrix elements of this supergroup arises. We note that
this problem is very complicated in the general case. In~[18], we could
evaluate the chiral representation, when matrix~(27) involves only the
Grassmann fields marked with an asterisk and bosonic fields. Below, we keep
to the following strategy.

We first evaluate the matrix consisting of only the components of the
electric field $E$ and the magnetic field $H$, whose {\sl contribution to
the effective functional is zero in the fermionic fields}:
\begin{align}
\Omega&=\exp\begin{pmatrix}E_z&0&0&E^+\\[1mm]
0&H&H^+&0\\[1mm]0&H^-&-H_z&0\\[1mm]E^-&0&0&-E_z\end{pmatrix}=
\nonumber
\\[2mm]
&=\begin{pmatrix}\ch E+E_z\dfrac{\sh E}E&0&0&E^+\dfrac{\sh E}E\\[1mm]
0&\ch H+H_z\dfrac{\sh H}H&H^+\dfrac{\sh H}H&0\\[1mm]
0&H^-\dfrac{\sh H}H&\ch H-H_z\dfrac{\sh H}H&0\\[1mm]
E^-\dfrac{\sh E}E&0&0&\ch E-E_z\dfrac{\sh E}E\end{pmatrix}.
\end{align}
The Cartan differential form is given by
\begin{equation}
\Omega^{-1}d\Omega=\begin{pmatrix}\omega_{11}&0&0&\omega_{14}\\[1mm]
0&\omega_{22}&\omega_{23}&0\\[1mm]0&\omega_{32}&\omega_{33}&0\\[1mm]
\omega_{41}&0&0&\omega_{44}\end{pmatrix}
\end{equation}
or
\begin{align}
\Omega^{-1}d\Omega={}&\omega_{11}X^{00}+\omega_{14}X^{02}+\omega_{41}X^{20}+
\omega_{44}X^{22}+\omega_{22}X^{++}+{}
\nonumber
\\[2mm]
&{}+\omega_{23}X^{+-}+\omega_{32}X^{-+}+\omega_{33}X^{--}.
\end{align}
We consider a three-dimensional curved space with the curvature radius
defined in terms of the invariant $E$. A similar argument can also be
presented for the $H$ vector, which means their independence of the
coordinates; we therefore assume that $E=\text{const}$, $dE=0$, and
$H=\text{const}$, $dH=0$. This leads to the constraints for the components
of the $E$ and $H$ vectors
\begin{equation}
dE^2=2E_z\,dE_z+E^+dE^-+E^-dE^+=0,\qquad dH^2=0,
\end{equation}
once again indicating the curved nature of both the base space and the space
where the dynamic fields exist.

For convenience, we represent the Cartan 1-forms as matrix elements of two
2$\times$2 matrices: we have
\begin{equation}
\begin{pmatrix}\omega_{22}&\omega_{23}\\[1mm]
\omega_{32}&\omega_{33}\\[1mm]\end{pmatrix}=\begin{pmatrix}
g_1\dfrac{\sh H}HdH_z-\dfrac{\sh^2 H}{H^2}H^+dH^-&
\dfrac{\sh^2 H}{H^2}H^+dH_z+g_1\dfrac{\sh H}HdH^+\\[1mm]
-\dfrac{\sh^2 H}{H^2}H^-dH_z+g_2\dfrac{\sh H}HdH^+&
-g_2\dfrac{\sh H}HdH_z-\dfrac{\sh^2 H}{H^2}H^-dH^+\end{pmatrix},
\end{equation}
where
$$
dH_z=\frac{\ptl H_z}{\ptl x}dx+\frac{\ptl H_z}{\ptl y}dy+
\frac{\ptl H_z}{\ptl z}dz
$$
and similarly for $dH^\pm$, $g_2=\ch H-H_z\sh H/H$, and
$g_1=\ch H+H_z\sh H/H$. The matrix $\bigl(\begin{smallmatrix}
\omega_{11}&\omega_{14}\\[1mm]\omega_{41}&\omega_{44}\end{smallmatrix}\bigr)$ for
the electric field components is entirely equivalent to matrix~(32) with the
replacement $H_z\to E_z$, $H^\pm\to E^\pm$. In what follows, we use
expression~(32) to obtain the effective functional for the dynamic magnetic
fields.

The Cartan forms for the fermionic Hubbard generators are formed similarly
to the bosonic case, with the only difference that odd Grassmann fields and
the corresponding differentials are involved and the inverse matrix is
calculated differently. In the final analysis, these Cartan 1-forms yield a
functional of the fermionic dynamic fields.

\section{Effective functional}

The construction of a ``symmetry tower" was proposed in~[4]. It is based on
the following observation: the variables in strongly correlated systems,
being generators of some superalgebras under a variation of the Hubbard
repulsion and the chemical potential, constitute a chain of superalgebras
$(S2,S1,S0)$ that specify excitations of those phases that emerge as the
repulsion increases. This chain, called the ``symmetry tower," is
characterized by the number of odd generators of the superalgebras and
starts with a minimal superalgebra that has two fermionic generators of the
algebra of spinless fermions.

In this section, we proceed from this superalgebra and construct its
deformation, which is nonlinear in the (super)generators involved. We then
evaluate the nonlinear exponential representation of the obtained
superalgebra. Next, using formula~(1) with Hamiltonian~(3), we obtain the
effective functional for the fields taking values in the chosen supergroup.
We note that the quantum super-Yangian first occurred as an example of a
nonlinear superalgebra in the one-dimensional Hubbard model in~[19].

We take the expression for the supergroup yielding the supercoherent states
$|G\kt=U|0\kt$ for the specified regime observed in the Hubbard model under
strong repulsion and half-filling in the form
\begin{equation}
U=\exp\begin{pmatrix}E_z&\chi&\chi&E^+\\[1mm]
\chi^*&H_z&H^+&\chi\\[1mm]\chi^*&H^-&-H_z&\chi\\[1mm]
E^-&\chi^*&\chi^*&-E_z\end{pmatrix}.
\end{equation}
The generators ($c^+$, $c$, $c^+c$, and $1-\gamma_5/2$) involved here are
equal to the corresponding matrices at the dynamic fields in the exponent
and form a closed superalgebra of a nonlinear type. We give matrix
expressions for the relevant generators:
\begin{equation}
\begin{alignedat}2
&c^+=\begin{pmatrix}0&0&0&0\\[1mm]1&0&0&0\\[1mm]
1&0&0&0\\[1mm]0&1&1&0\end{pmatrix},&\qquad
&c=\begin{pmatrix}0&1&1&0\\[1mm]0&0&0&1\\[1mm]
0&0&0&1\\[1mm]0&0&0&0\end{pmatrix},
\\[2mm]
&c^+c=\begin{pmatrix}0&0&0&0\\[1mm]
0&1&1&0\\[1mm]0&1&1&0\\[1mm]0&0&0&2\end{pmatrix},&\qquad
&\frac{1-\gamma_5}2=\begin{pmatrix}0&0&0&0\\[1mm]
0&1&0&0\\[1mm]0&0&1&0\\[1mm]0&0&0&0\end{pmatrix}.
\end{alignedat}
\end{equation}
The (anti)commutation relations are
\begin{equation}
\begin{aligned}
&\{c^+,c\}_+=2(1+s^++s^-),
\\[2mm]
&[c^+,s^+]_-=c^+\frac{1-\gamma_5}2\frac{1-s_z}2-
\frac{1-\gamma_5}2\frac{1+s_z}2c^+,
\\[2mm]
&[c^+,s^-]_-=c^+\frac{1-\gamma_5}2\frac{1+s_z}2-
\frac{1-\gamma_5}2\frac{1-s_z}2c^+,
\\[2mm]
&[c,s^+]_-=c\frac{1-\gamma_5}2\frac{1-s_z}2-
\frac{1-\gamma_5}2\frac{1+s_z}2c,
\\[2mm]
&[c,s^-]_-=c\frac{1-\gamma_5}2\frac{1+s_z}2-
\frac{1-\gamma_5}2\frac{1-s_z}2c,
\\[2mm]
&[c^+c,s^+]_-=-s_z,\qquad[c^+c,s^-]_-=s_z,\qquad[c^+c,s_z]_-=s^--s^+,
\\[2mm]
&[c^+,\rho^+]=c\frac{1-\gamma_5}2-\frac{1-\gamma_5}2c,\qquad
[c^+,\rho^-]=0,\qquad[c^+,\rho_z]=c^+,
\\[2mm]
&[c,\rho^+]=0,\qquad[{c,\rho^-}]=c\frac{1+\gamma_5}2-
\frac{1+\gamma_5}2c,\qquad[c,\rho_z]=-c,
\\[2mm]
&[c^+c,\rho^+]=-2\rho^+,\qquad[c^+c,\rho^-]=-2\rho^-,\qquad[c^+c,\rho_z]=0.
\end{aligned}
\end{equation}
We do not take the coefficients of the matrix representation of
operators~(15) and~(17) into account here.

Calculating a nonlinear representation of the above (super)group is
cumbersome and quite laborious. The problem of calculating the inverse
(super)matrix, which is required for calculating the Cartan
(super)differential forms, is especially complicated. In the calculations,
we used a computer with 2\,Gbytes of RAM and the Mathematica~7.0 symbolic
calculation system extended by the superEDC program~[20] for working with
Grassmann numbers, (super)matrices, and Cartan (super)differential forms.
The (super)matrix elements for the matrix $U$ in (33) and the Cartan 1-forms
were evaluated in~[21], where all the notation used here was also given.
With the stabilized algorithm for working with superfields that take values
in (super)matrices, the total time for computing the effective functional
given below was about 50~hours.

In calculating functional~(1), we used the expansion of the kinetic energy
in formula~(3) and a representation of the product of operators at the
neighboring sites in the form of a series in the lattice constant. In the
first order, we obtain
\begin{align}
a(j)a^+(j+b)&=a(j)\exp\biggl(b\frac{\ptl}{\ptl x}\biggr)a^+(j)=
\nonumber
\\[2mm]
&=a(j)\biggl[1+(b\cdot\nabla)+
\frac{(b\cdot\nabla)^2}{2!}+\cdots\biggr]a^+(j)=
\nonumber
\\[2mm]
&=a(x)a^+(x)+a(x)(b\cdot\nabla)a^+(x)+\ldots.
\end{align}

We replace the index $j$ with the coordinate $x=bj$ for future use in
deriving the representation of the effective functional. We note that the
operators $a$ and $a^+$ are expressed in terms of the Hubbard generators.

Expanding the exponential in a series using formula~(36) and the expression
for the Cartan (super)differentials in the first order in the lattice
constant, we obtain the effective fermionic functional in the form of the
sum
\begin{equation}
\hat L_1=(k_1+k_2)\chi(r)(b\cdot\nabla)\chi^*(r)+k_3\chi(r)\chi^*(r).
\end{equation}
This functional is quadratic in the spinless fermionic field and contains
the dynamic fields $H_i$ and a coordinate dependence that gives the metric
of the curved space. The coefficients $k_i$ in~(37) are given by
\begin{equation}
\begin{aligned}
&k_0=EH[f_3+f_2h_3+f_2(x+z)],
\\[2mm]
&k_1=f'_2+f'_{2h}+2f'_{1hh}(h_1+x+y+z)+f'_{hh}[H_zh_2+zh_2+(x+y)(H_z+z)],
\\[2mm]
&k_2=k_0[E^2(f_3-f_2h_4)+(y+z)(zf_3-E^2f_2)+(H_z-z)(f_4-f_3h_4)],
\\[2mm]
&k_3=k_0\{2(f_4-h_4 f)+(x-y-h_5)(f_3-h_4f_2)-(y+z)[2f_3+(x-y)f_2]\}.
\end{aligned}
\end{equation}
These expression include contributions from the derivatives of the magnetic
fields; the exact formula for them will be given elsewhere. The other
elements of the condensed notation are presented in the appendix.

Matrix~(32) gives differentials of the dynamic fields. We can see from~(1)
which derivatives of these fields appear. It follows from~(1) that the
effective functional involves only the expression for $\omega_{22}$. With
all the foregoing, we obtain the expression for the effective functional of
bosonic fields:
\begin{equation}
\hat L_0=-[H^-(b\cdot\nabla H^+)+H^+(b\cdot\nabla H^-)+
H_z(b\cdot\nabla H_z)]\frac{\sh^2H}{H^2}+H_z(b\cdot\nabla H_z)
\frac{\sh H}H\ch H.
\end{equation}
If the second-order terms in the lattice constant are kept in the kinetic
energy expansion in~(36), then an effective functional can be obtained that
contains quadratic contributions from derivatives of the magnetic fields.
This then gives a nonlinear sigma-model, which is an analogue of the
Heisenberg model in the functional integral for the Hubbard model. A
functional linear in the derivatives is similar to the functional in the
Chern--Simons model, which is natural for our problem.

Finally, the full effective functional for the spinless version of the
Hubbard model has the form
\begin{equation}
\hat L_\mathrm{eff}=\hat L_0+\hat L_1.
\end{equation}
We believe that the regime we have chosen in the Hubbard model is related to
the spin-liquid phase revealed in the organic metal
(BEDT--TTF)$_2$Cu$_2$(CN)$_3$. This compound requires taking a
two-dimensional lattice with a hexagonal symmetry.

\section{Conclusions}

We have offered a further development of the approach formulated
in~[3],~[4]. We showed that the supercoherent state of form~(33) defines a
nonlinear representation of the minimal (super)extension of the Lorentz
group. Continuing this procedure of (super)extension leads to a
representation of the superconformal group in the superspinor basis,
nonlinear in the fermionic fields. We constructed a new nonlinear matrix
representation for the translation generators in the $(2{+}1)$-dimensional
space--time. It defines a nonlinear superalgebra that enlarges the set of
(super)algebras that have been proposed in the ``symmetry tower." We note
that after~[19], the search for a quantum algebra in the Hubbard model was
conducted very vigorously but unfortunately unsuccessfully. We have
indicated a way to introduce similar structures in the Hubbard model.
Superalgebra~(35) contains pairs of odd generators that anticommute with the
shift operator (for a spherical space, a shift is equivalent to a rotation),
as is appropriate for the standard supersymmetry algebra.

The effective functional of form~(40) explicitly contains the coordinates of
the $SO(2,1)$ or $SO(3)$ group spaces and describes the dynamics of Bose and
Fermi fields in a three-dimensional curved space--time. A functional of this
type, as far as we know, was first obtained in the Hubbard model.

In relation to this work, we note an entirely new problem for the Hubbard
model: calculating the cohomology groups, the second and the third in
particular, with the purpose of studying the superalgebras in the ``symmetry
tower." This approach opens the possibility of studying the mechanism of the
Mott--Hubbard-type metal--insulator transition as a spontaneous violation of
the (super)conformal (super)group.

\section*{Appendix}

The additional notation in formulas~(38) is
\begin{align*}
&f(\alpha)=\frac{\bigl(\sh(\alpha E)\bigr)/E-\bigl(\sh(\alpha H)\bigr)/H}
{E^2-H^2},\qquad f_{hh}=\frac{\ptl}{\ptl(H^2)}[H^2f(\alpha)]_{\alpha=1},
\\[2mm]
&f_{1hh}=\frac{\ptl}{\ptl\alpha}(f_{hh})_{\alpha=1},\qquad
f_{2h}=\frac{\ptl}{\ptl H}(Hf_2),
\\[2mm]
&f_2=\frac{(\sh E)/E-(\sh H)/H}{E^2-H^2},\qquad
f_3=\frac{\ch E-\ch H}{E^2-H^2},\qquad f_4=\frac{E\sh E-H\sh H}{E^2-H^2},
\\[2mm]
&f_{hh}'=\frac{\sh H}Hf_{hh},\qquad f_{1hh}'=\frac{\sh H}Hf_{1hh},\qquad
f_2'=\frac{\sh H}Hf_2,\qquad f'_{2h}=\frac{\sh H}Hf_{2h},
\\[2mm]
&\begin{alignedat}3
&h_0=H^+-H^-,&\qquad&h_1=H^++H^-,&\qquad&h_2=H^-+H_z,
\\[2mm]
&h_3=H^--H_z,&\qquad&h_4=H^++H_z,&\qquad&h_5=H^+-H_z.
\end{alignedat}
\end{align*}

\subsection*{Acknowledgments}
One of the authors (V.~M.~Zh.)~is grateful to V.~V.~Kiselev and
A.~B.~Borisov for the discussion on the problems touched upon here.

\end{document}